\documentclass[iop,apjl]{emulateapj}
\usepackage{natbib}
\usepackage{graphicx}
\begin{document}

\newcommand{\kms}{\ensuremath{\mathrm{km}\,\mathrm{s}^{-1}}}
\newcommand{\MLsun}{\ensuremath{\mathrm{M}_{\sun}/\mathrm{L}_{\sun}}}
\newcommand{\Lsun}{\ensuremath{\mathrm{L}_{\sun}}}
\newcommand{\Msun}{\ensuremath{\mathrm{M}_{\sun}}}
\newcommand{\Aunits}{\ensuremath{\mathrm{M}_{\sun}\,\mathrm{km}^{-4}\,\mathrm{s}^{4}}}
\newcommand{\surfdens}{\ensuremath{\mathrm{M}_{\sun}\,\mathrm{pc}^{-2}}}
\newcommand{\voldens}{\ensuremath{\mathrm{M}_{\sun}\,\mathrm{pc}^{-3}}}
\newcommand{\gevcc}{\ensuremath{\mathrm{GeV}\,\mathrm{cm}^{-3}}}
\newcommand{\etal}{et al.}
\newcommand{\LCDM}{$\Lambda$CDM}
\newcommand{\ML}{\ensuremath{\Upsilon_*}}

\title{Missing the Point --- a Brief Reply to Foreman \& Scott and Gnedin}

\author{Stacy S. McGaugh}
\affil{Department of Astronomy, University of Maryland,
College Park, MD 20742-2421, USA}

\begin{abstract}
In recent postings, \citet{FS11} and \citet{gnedin11} criticize my work on the baryonic
Tully-Fisher relation (BTFR) of gas rich galaxies as tests of MOND and \LCDM\ \citep{BTFRmond,btfseb}.  
These criticisms are rather redundant, as they mostly rehash material I have already discussed.
\citet{gnedin11} is concerned with explaining away the problem of the apparently missing baryons with 
ionized gas.  This is one hypothetical 
possibility that I have previously discussed \citep{M10,MWolf,btfseb}.  \citet{FS11} claim to find a difference between
the acceleration scale that they measure and that I measure from the same data, but the apparent difference 
results simply from their failure to account for the well known fact \citep{BT} that flattened disks rotate faster than the 
equivalent spherical mass distribution \citep[as spelled out for this application in][]{btfseb}.
They further argue that the the intrinsic scatter in the BTFR is merely tiny rather than zero.  
This difference is within the uncertainty of the uncertainties, so confirms rather than refutes my result.
Worse, these papers miss the basic point.  Why does MOND have \textit{any} predictions come true?  
Why does it provide such an effective and economical description of galaxy dynamics?  
They do nothing to address these deeper questions.
One can imagine that the apparently MONDian behavior of galaxies 
is merely an approximation to some phenomenology that emerges
in the context of \LCDM, but we have yet to demonstrate how this occurs.
Until we confront this issue seriously, and confirm the existence of non-baryonic cold dark matter
in the laboratory, it is reasonable to remain skeptical of \LCDM\ as well as of MOND. 
\end{abstract}

\keywords{dark matter --- galaxies: kinematics and dynamics --- galaxies: dwarf}

\section{Introduction}

In the present cosmological paradigm, \LCDM, most of the mass-energy of the universe is in invisible
components: dark matter and dark energy.  Generically speaking, dark matter could mean any non-luminous
gravitating mass, baryonic or not.  For our cosmological model to work, the dark matter must specifically be dynamically
cold and non-baryonic.  

We have arrived at the \LCDM\ paradigm for very good reasons that do not require repetition here.
However, to date all evidence for dark matter remains astronomical in nature,
resting fundamentally on the assumption that no new gravitational physics occurs when extrapolating 
the force law by eight orders of magnitude from precision solar system tests to extragalactic systems. 
The persistent presence of mass discrepancies in extragalactic systems might therefore also be interpreted
as a failure of this assumption.

We come then to a fundamental dichotomy of attitudes that different scientists bring to the problem.
A common one is that \LCDM\ has so many successes that it \textit{must} be correct.  Consequently,
non-baryonic cold dark matter \textit{must} exist.  This is the attitude I brought to the problem.
However, another valid attitude is the \LCDM\ \textit{requires} a hypothetical new particle generally
presumed to exist in a hypothetical new dark sector of particle physics.  Until
the existence of these particles is confirmed in the laboratory, \LCDM\ remains an unconfirmed
hypothesis built on a good but unproven supposition.  
Which of these attitudes one brings to the problem inevitably colors one's interpretation of 
any given set of data.

\section{Some Context --- My Experience with the Tully-Fisher Relation}

Rotating galaxies obey a relation between their rotation velocity and luminosity \citep{TForig}, or more generally, 
their baryonic mass \citep{btforig} of the form
\begin{equation}
M_b \propto V^x
\end{equation}
with the slope typically found to be in the range $3 \le x \le 4$.
Though widely used in distance scale work, the physical basis of this relation remains unclear.  
Starting from Newton's universal law of gravity, \citet{aaronson} obtained
\begin{equation}
\Sigma M \propto V^4
\end{equation}
where $\Sigma$ is the mass surface density.  This suffices if, as believed at the time,
disk galaxies all share essentially the same surface brightness \citep{freeman70} so that
$\Sigma \sim$ constant.

I first encountered this problem in the context of low surface brightness (LSB) galaxies, which explicitly violate the
required constancy of $\Sigma$.  I had, at the time, developed my own (perfectly conventional) theory for the
formation of LSB galaxies.  Motivated by observations of the stellar populations of LSB galaxies \citep{LSBpops}
that suggested that LSB galaxies were less evolved and perhaps had a later start in life than their higher
surface brightness brethren, I hypothesized that their low stellar densities followed from residing in low density 
dark matter halos.  Though I arrived at this idea independently, a similar idea had been previously discussed by
\citet{DS}.

In modern parlance, I associated low surface brightness galaxies with late forming dark matter halos.
This hypothesis made two predictions: (1) that LSB galaxies should be less strongly clustered than HSB galaxies,
and (2) that LSB galaxies should be systematically shifted off of the Tully-Fisher relation.  The first of these
hypotheses was confirmed \citep{MMB94} and subsequently reconfirmed \citep{rosenbaum04,rosenbaum09}.
I therefore engaged briefly in the (not uncommon) fantasy that I understood galaxy formation.

The data soon falsified prediction 2:  LSB galaxies fall on the same Tully-Fisher relation with the same normalization
as higher surface brightness galaxies \citep{zwaanTF,sprayTF}.  Shocked, I struggled hard to understand how this could
be, without success \citep{MdB98a}.  Every time I thought I had succeeded, I realized that at some point I had made an assumption
that guaranteed the desired result.  Since that time, various papers have been published that assert success but appear
to suffer from the same shortcoming.

Though shocked enough at the failure of my own prediction and the difficulty of recovering a conventional model, it was
more shocking still to realize that this observation had been predicted in advance by \citet{milgrom83}.
\textit{A priori} predictions play an important role in the philosophy of science because it is often possible 
to ``fix'' a model once a result is known\footnote{Remarkably, Milgrom's prediction was made at a time when 
LSB galaxies were widely thought not to exist because of Freeman's Law.}.  Dark matter
models are certainly flexible enough to accommodate such fixes.  

Were it permissible to simply ignore Milgrom's prediction, then I would have patched up my hypothesis and claimed
success.  This would have been convenient, emotionally fulfilling, positive for my career, but intellectually dishonest.  
I did not get it right \textit{a priori}.  MOND did.  I relate this story because I value intellectual honesty very highly.  
I am willing to admit when I am wrong, and have done so in the past as I did with my prediction for the Tully-Fisher
relation of LSB galaxies.  The current situation with the gas rich galaxies is not such an occasion.

\section{Specific Criticisms}

\citet{gnedin11} and \citet{FS11} make distinct criticisms of \citet{BTFRmond}.  
\citet{gnedin11} focuses on potentially unobserved baryons while \citet{FS11} focus more on the observed scatter
in the Tully-Fisher relation.  I can find little \textit{important, factual} information in either paper that I disagree with.
Indeed, I find myself in the bizarre position of defending myself against points they make as if they were new when
I have already made many of the same points myself \citep[e.g.,][]{M10,btfseb}.

Both seem to misconstrue the meaning of the starting equation typically invoked for discussing the BTFR in the context of \LCDM:
\begin{equation}
M_{\Delta} = C_{\Delta} V_{\Delta}^3,
\end{equation}
where $C_{\Delta}$ is an easily derived scaling constant that depends on the choice of over-density $\Delta$.
As I discuss in \S 3.1 of \citet{btfseb}, there does at least appear to be a consensus on this relation, which refers
to the \textit{total} mass enclosed by $\Delta$ and the circular velocity of a test particle at the radius enclosing this mass.
This is not the observed BTFR.  

An obvious and natural route to obtaining a first approximation of the BTFR is to
define factors $f_d$ and $f_V$ that relate the observed fraction of baryons to the baryonic mass associated with the
dark matter halo ($M_b = f_d f_b M_{\Delta}$) and the observed velocity to $V_{\Delta}$ ($V_f = f_V V_{\Delta}$).
While these fudge factors are presumably of order unity, they do not have to be identically so.

This is a standard approach adopted, for example, by \citet{MMW98}.  The only difference is a trivial one of definition in $f_d$:
they choose $\Delta = 200$ and subsume the universal baryon fraction $f_b$ into into their parameter $m_d = M_b/M_{200}$.  
They \textit{assumed} $m_d$ was constant, yielding a BTFR with slope 3.  This fit the data for the bright galaxies known
at the time tolerably well with $m_d \approx 0.05$, somewhat less than $f_b = 0.17$.  This is perfectly admissible, though it
does raise the question that \citet{gnedin11} attempts to address:  where are those unobserved baryons?

The new data for gas rich galaxies implies a much lower $m_d \approx 0.01$ \citep{gurov}, as illustrated in Fig.~4 of
\citet{btfseb}.  In order to match the data, we need to make the combination of $f_d$ and $f_V$ vary with velocity as
defined in equations 16 and 17 of \citet{btfseb}.  That is an inference of what is required, nothing more.  
Any model that matches this constraint successfully passes this particular test.

\begin{figure}[h]
  \centerline{\includegraphics[width=14.5cm]{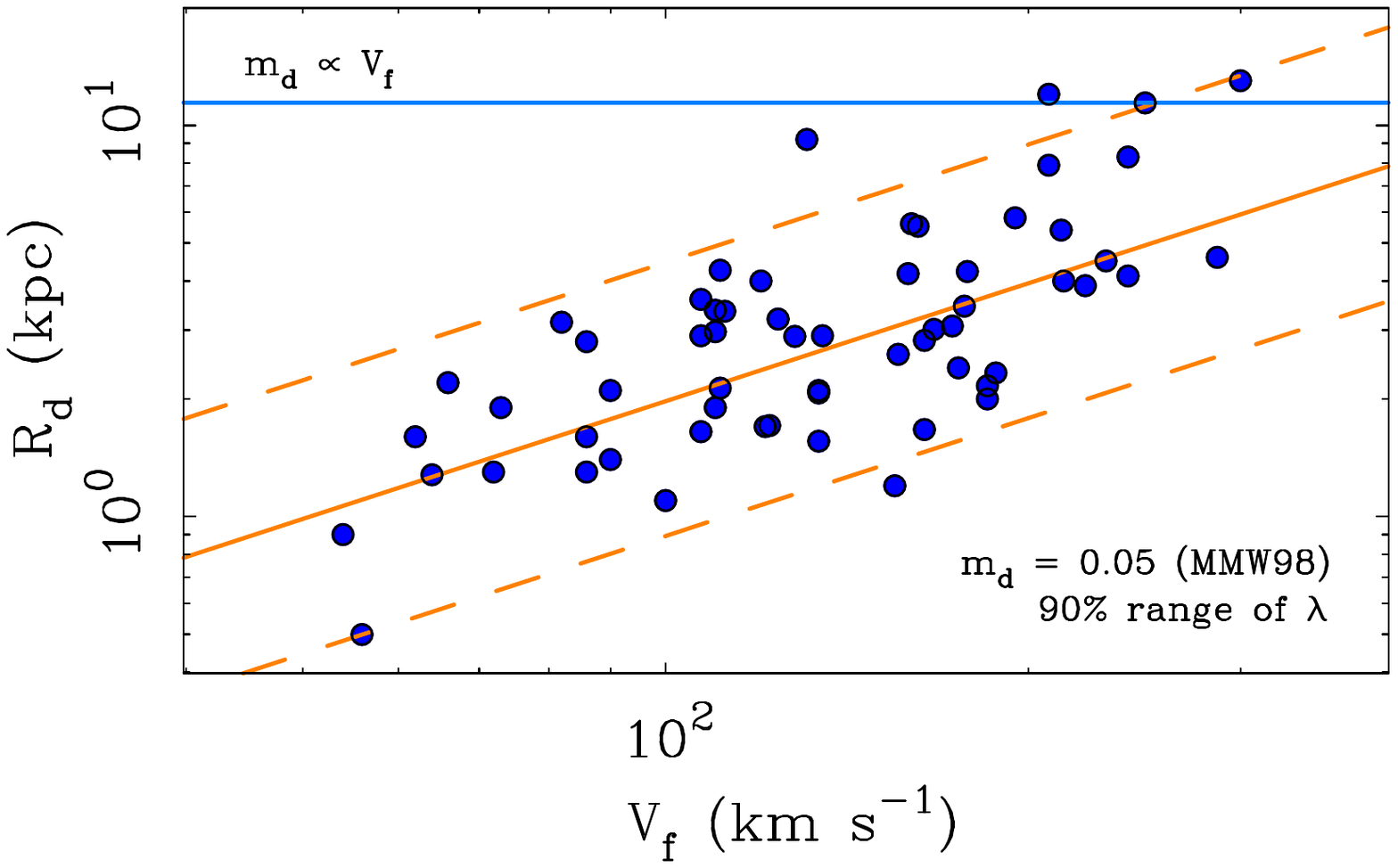}}
  \caption{The disk scale length--rotation velocity relation \citep[analogous to Fig. 4 of][]{MMW98}.  
  Models of they type \citet{MMW98} discuss do a good job of explaining the data.
  Both the mean trend and the scatter are well matched by a constant baryon fraction $m_d$ (solid orange line)
  with the range of spin parameters $\lambda$ expected from numerical simulations (dashed orange lines).  
  However, in order to fit the BTFR, one needs to drop the assumption of constant $m_d$ and 
  scale it as $m_d \propto V_f$.  This fits the BTFR, but destroys the consistency with the
  scale length--rotation velocity relation (horizontal line.  The normalization of this line is adjustable but
  not the slope.)  This illustrates the general principle that one must be careful that tuning
  done to fit one observation does not upset consistency with another.
  }
  \label{VRd}
\end{figure}

Indeed, I constructed such a model myself over a decade ago.  Starting from the model of \citet{MMW98}, I noted that
the BTFR could be fit by making $m_d$ a linear function of circular velocity.  I never published this model because it is
obviously wrong.  Though one can indeed fit the BTFR by fine-tuning $m_d \propto V_f$, this destroys other virtues of
the model.  For example, the disks size--circular velocity relation, which is well explained with constant $m_d$, now fails
(Fig.~\ref{VRd}) because the dependence of disk scale length $R_d$ on $m_d$ and circular velocity in their equation~12
cancel out.

I have spent a great deal of time over the past 17 years attempting to construct acceptable \LCDM\ models.  I have not
succeeded.  There are no lack of those who have claimed success, but these models are rarely consistent with one another.
At most one of these models can be correct, so it seems more likely that the claims of success are overstated
\citep[e.g.,][]{TGKPR}

\subsection{Gnedin}

\begin{figure}[h]
  \centerline{\includegraphics[width=14cm]{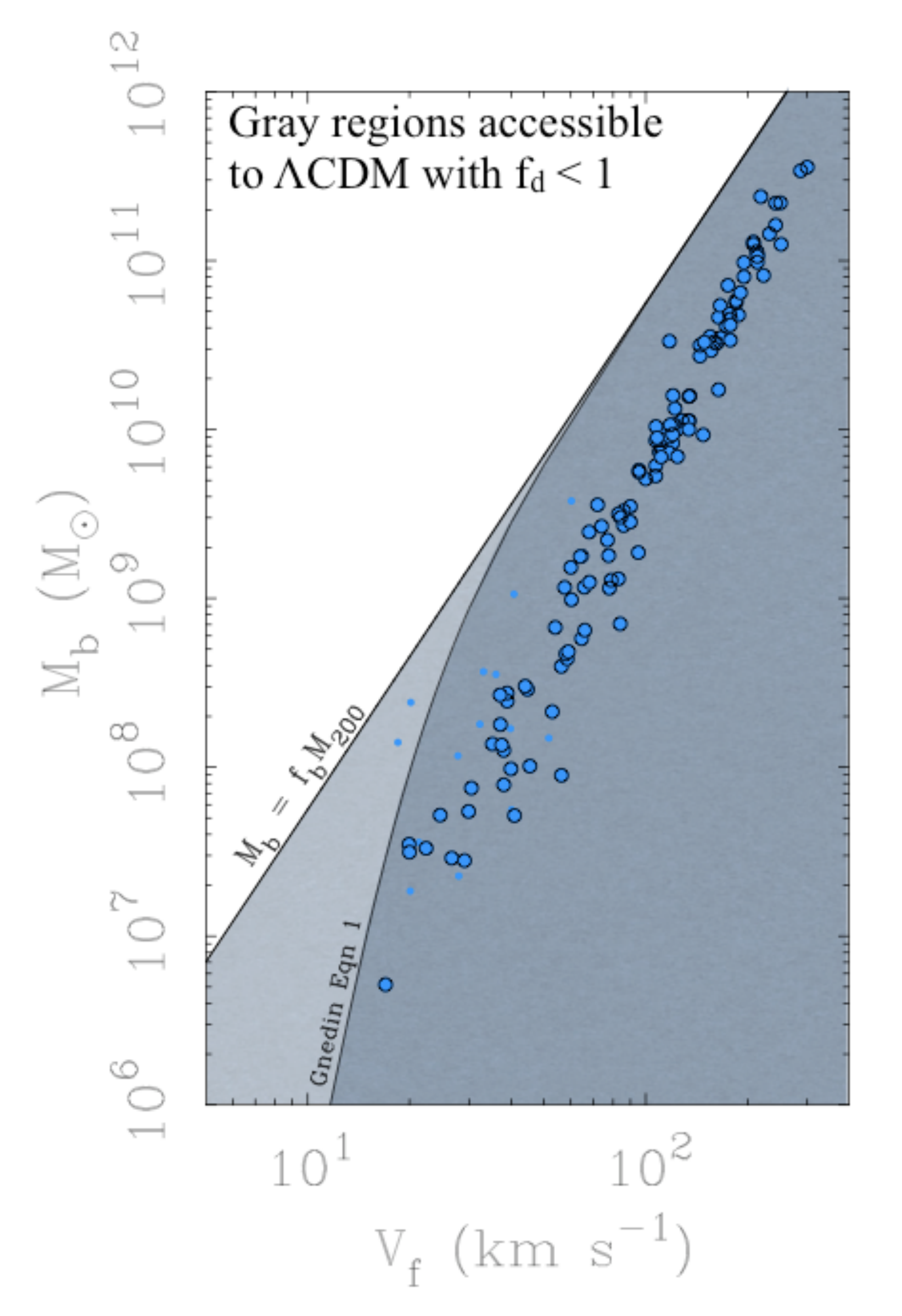}}
  \caption{The BTFR with all data discussed by \citet{btfseb}, making no distinction between gas and star dominated 
  galaxies.  Galaxies identified there as potentially having large
  systematic inclination errors are shown as smaller symbols.  Inclusion of these objects will increase the scatter
  and $\chi^2$, and may also skew the slope.  Also shown is the \LCDM\ mass--rotation velocity relation (straight
  line).  This shows all of the baryons enclosed within the overdensity $\Delta = 200$.  \LCDM\ does not require galaxies
  to fall on this line, though that would have been a very satisfactory result.  If not all baryons are detected ($f_d < 1$),
  galaxies may fall anywhere in the shaded region.  That they fall along such a narrow track when so much parameter
  space is available strikes me as a fine tuning problem.  This is only compounded by the fact that the data follow the
  particular BTFR that is unique to MOND.  The dark gray region represents the allowed regime according to
  \citet{gnedin11} --- the curved line is his  equation 1 with $\alpha = 1$.  Reionization may well reduce the cold baryons available to be
  detected in low mass halos, but by itself it does nothing to explain the observed BTFR.  Note that if \citet{gnedin11} is correct
  that this is an upper limit to the baryon content of galaxies in \LCDM, the two galaxies left of the curved line 
  (which I independently identified as having very uncertain inclinations) are impossible in \LCDM.  
  Indeed, the line moves right and through the data as $\alpha$ increases, so  $\alpha > 1$ can be excluded
  \citep[see also][]{tikhonov,MWolf}.
  }
  \label{TFshaded}
\end{figure}

\citet{gnedin11} does not claim to present a model that matches the data.  
He merely argues that plausible astrophysical processes will reduce $f_d$ below unity.
I agree entirely that this is possible (Fig.~\ref{TFshaded}).  
If so, it is incumbent upon us to find and identify those baryons, not merely accept that
they are there is some unobserved form\footnote{The \LCDM\ missing baryon problem is not subtle.  For every detected
baryon in a $\sim 20\;\kms$ galaxy, $\sim 50$ remain undetected \citep{M10}.  Much of the community seems to accept
this without blinking.  Yet when MOND suffers a factor of $\sim 2$ missing baryon problem --- as it does in clusters of galaxies
\citep{SMmond} --- we are quick to believe it is falsified.}
(in this case, ionized gas).  This was explicitly discussed in \citet{M10}.  
Indeed, \citet{MWolf} have already published (their Fig.~4) a version of Gnedin's Fig.~2 that shows that re-ionization alone
is inadequate to explain the variation of $f_d$.  This is one example of what I mean to find it bizarre to address as criticisms 
points I have already published myself.

\subsection{Foreman \& Scott}

The criticism of \cite{FS11} is considerably less collegial in tone.  They appear to be suffering from an ancient emotional impulse:  
when you don't like the message, kill the messenger.  Anyone who has attempted to discuss the evidence for 
evolution or global warming in a public setting has no doubt experienced this impulse first hand.

A classic approach that is employed when one wants to belittle an argument is to focus on a tiny piece of it, show it is
wrong, and thus cast doubt on the entire work.  This is the approach \citet{FS11} adopt.  They analyze the
the same data as myself\footnote{www.astro.umd.edu/\~{}ssm/data/gasrichdatatable.txt} and
find that the peak $V_f^4/GM_b = 1.59\;\textrm{\AA}\,\mathrm{s}^{-2}$ is different from the 
$a_0 = 1.3\;\textrm{\AA}\,\mathrm{s}^{-2}$ that I find \citep[see Table 2 of][]{btfseb}.  This is not some horrible mistake
as they would apparently like to believe.  As discussed in \S 3.2.2 of \citet{btfseb}, a paper they cite but seem
not to have read, the acceleration scale estimated in this
manner is identical to the MOND acceleration scale only in the limit of infinite distance from an isolated point mass.
Since such a condition in never obtained in reality, there is a correction factor of order unity that follows simply form the fact
that a flattened mass distribution like a disk galaxy spins faster than the equivalent spherical mass distribution.  
This is a basic result that should be familiar to any student of galactic dynamics \citep{BT}.
This factor (0.8) is the difference between the acceleration scale of \citet{FS11} and myself.  

\citet{FS11} also argue that the scatter about the BTFR is slightly larger than I find.
While I certainly consider the problem of the missing baryons to be a puzzle, it is at least approachable.  
The greater problem for \LCDM, to my mind, is the small intrinsic scatter.  Once we allow for 
undetected baryons ($f_d < 1$), as we must in order to reconcile the \LCDM\ halo mass--rotation velocity relation with the
observed BTFR, then \textit{any} part of the Tully-Fisher plane below the line $M_b = f_b C_{\Delta} V_{\Delta}^3$  is accessible to 
\LCDM\ models (Fig.~\ref{TFshaded}).  
It is easy to conceive of models where galaxies of a given circular velocity have $f_d$ ranging from near unity
to near zero.  Indeed, it is not uncommon to see such a range in numerical simulations.  The scatter of the resultant BTFR
would be measured in dex rather than tenths of a dex.  So the real question is why the scatter is so small. 
How do galaxies ``know'' to accrete just the right fraction of cold baryons for their rotation velocity?

Though I am hardly the first to find negligible scatter in the Tully-Fisher relation \citep[e.g.,][]{verhTF,TFscatterBersh,trach},
\citet{FS11} make the case that there is some scatter in the data I used: 0.28 dex rather than the 0.24 dex expected purely from
the median observational error.  Taking this at face value, and the difference in quadrature, leaves an intrinsic scatter of 0.15 dex ---
precisely the value I give in \citet{btfseb}.  They also analyze a larger sample that has a somewhat larger scatter ---
0.33 dex.  If the typical errors are no larger, this implies an intrinsic scatter of 0.22 dex: still a remarkably small value given
the potential variation in $f_d$ alone \citep[never mind the many other potential sources of scatter, e.g.,][]{EL96}.

I do not know the contents of the larger sample \citet{FS11}, 
so I cannot repeat their analysis.  However, if it is indeed composed of gas
rich galaxies, I would surmise that a goodly number of the additional galaxies are likely to come from the data of \citet{begum}.
These galaxies are wonderful for being some of the most extreme (in terms of low velocity) yet observed.  They are also
very difficult observational targets.  One of the harder parameters to adequately constrain is the inclination, which can easily
lead to \textit{systematic} uncertainties that invalidate the usual assumptions of data analysis.  As discussed at considerable
length in \citet{btfseb}, I have analyzed only galaxies for which I am reasonably confident of the inclination estimate, as
indicated by broad agreement between optical and HI data.  If one includes all of these data, the scatter certainly goes up.
The best-fit slope also becomes somewhat shallower \citep{btfseb}, as these data skew to low velocity for a given mass.  
This skew is the expected signature of systematic inclination errors, as even a face-on galaxy may not appear perfectly circular if it
is affected by an oval distortion or bar.  It is hard to know for certain whether this is what is going on, but it is hardly surprising
that the scatter goes up incrementally as progressively less reliable data are included.

Dickering about whether the intrinsic scatter is zero or $\sim 0.2$ dex is of little consequence when variation in $f_d$ could easily
lead to an order of magnitude larger scatter.  
As it is, we have no clear prediction for the BTFR in \LCDM, let alone the scatter therein.  
One can certainly build models with little
scatter, but one must be careful (as I have repeatedly found) that we do not make an assumption that builds-in
the desired result.  Historically, this has routinely been done by simply assuming a constant value for $f_d$.

One common approach to explain $f_d < 1$ is to invoke feedback from supernovae.
This is not well understood and can have many different implementations.
The question then becomes how well these implementations represent reality.
I would expect a messy process like supernova feedback to result in considerable scatter --- 
a very small gas rich galaxy might experience zero, one, or numerous supernovae.
Such differences should propagate into scatter in the BTFR, where
the limited scatter budget can be consumed very rapidly. 

\citet{FS11} also claim to find higher $\chi^2$ than I do.  
I do not see an error in my code.  I do notice that if I fail to account for the slope of the
relation in the computation of the weights on the bivariate errors, then 
I get the same number for $\chi^2$ that they get for the same sample.

For the sake of argument, let us suppose that the higher $\chi_{\nu}^2 \approx 1.5$ is correct.  Now suppose that it
is \LCDM\ rather than MOND that predicts a relation with no scatter.  Do we claim that this falsifies \LCDM?  
Or do we report it as a success for the paradigm, with maybe a hint that there is some small contribution to the error 
budget that hasn't been entirely accounted for?

I have gone through the statistics of these data, and the various
subsamples, in excruciating detail \citep{btfseb}.  Doing so again would not prove anything, so I make use of an entirely
independent program written by Ben Weiner \citep{benfit}.  Treating all parameters as free gives a best fit slope of
$3.82 \pm 0.22$ with $\chi^2 = 33.5$ for 47 galaxies.  This slope is consistent with what I find myself for the same fit parameters
for the same data, but $\chi^2$ is somewhat lower than I found.  
Fixing the normalization and slope to the MOND values yields $\chi^2 = 58.5$ --- slightly more than I found but less than
\citet{FS11} find.  While it is annoying that there should be even a slight disagreement in such a simple calculation, the
difference is within the uncertainty of the uncertainties.  

\begin{figure}[h]
  \centerline{\includegraphics[width=14.5cm]{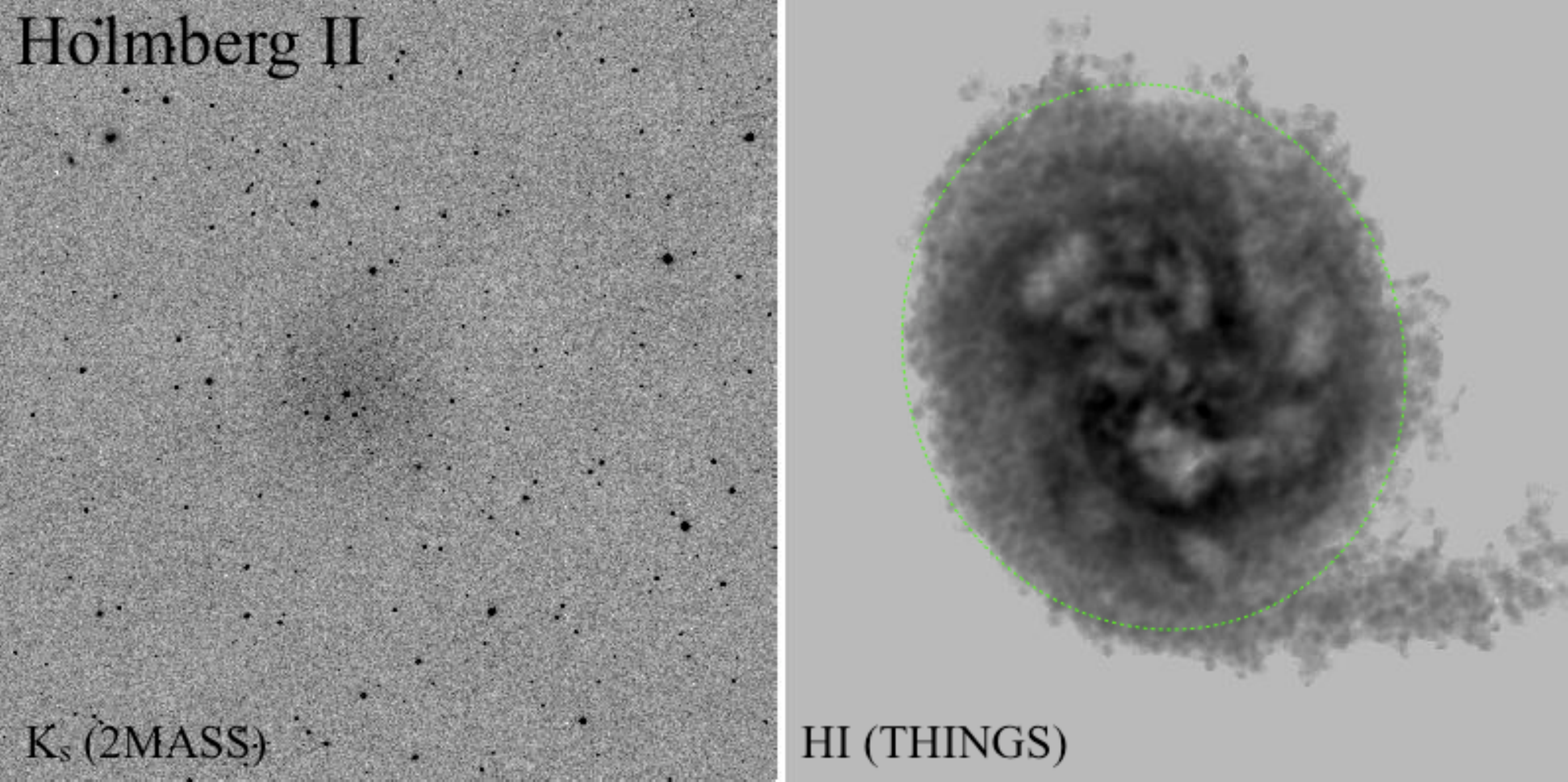}}
  \caption{The dwarf galaxy Holmberg II in stars as traced by the $K_s$ image from 2MASS \citep[left]{2MASS} and 
  atomic gas from THINGS \citep[right]{FTHINGS}.  The inclination of this galaxy is listed variously as $31^{\circ}$
  or $45^{\circ}$ \citep{THINGS} or more (see text).  I have refrained from utilizing data for cases where the inclination is this unsettled
  \citep{btfseb}; the larger scatter and $\chi^2$ \citet{FS11} find probably stems from including such cases.  
  In this particular galaxy, larger inclinations pose a problem for MOND \citep{hoii_sanchez}.  Consistency with the BTFR
  requires $i \approx 25^{\circ}$ (green ellipse).  That is, to believe that this galaxy poses a challenge to MOND
  requires you to believe that the inclination of this galaxy is significantly different from that illustrated by the ellipse,
  which is not a fit, just the projection of a circle inclined by $25^{\circ}$.  Note that masses in MOND scales as
  $[V_{obs}/\sin(i)]^4$, so even a tiny error in inclination can lead to a large error in mass for low inclination galaxies.
  }
  \label{HoII}
\end{figure}

It is a well known fact that in astronomy, some data are more reliable than others.  Rather than expanding the sample
as \citet{FS11} do, one needs, if anything, to restrict it further.  Doing so essentially returns us to the results of
\citet{stark}, who held the data to the highest standards of quality control.  Their empirical result is consistent with 
MOND\footnote{This is a good example of wanting --- or in this case, not wanting --- a particular result.  I did not want
to find a slope of 4 when we did this experiment because I knew it would eventually lead to exactly this sort of dispute.
The data didn't give us liberty to say otherwise, however.} ($x = 3.94 \pm 0.07$).  In trying to extend the sample, I have
included some, but not all, of the data from \citet{begum}.  These are difficult observational targets that are inevitably less
reliable than the data discussed by \citet{stark}.  Indeed, only three of the twenty-nine galaxies of \citet{begum} meet the
quality requirements of \citet{stark}.  As discussed in \citet{btfseb}, I reject about half the sample as having unreliable
inclinations.  An example of just how difficult it can be to measure the inclination of these objects is given by the case
of Holmberg II  (Fig.~\ref{HoII}), which has rather better data than the majority of comparably low mass galaxies.
However, inclination estimates for it vary from $31^{\circ}$ to $45^{\circ}$ \citep{THINGS} to $55^{\circ}$ or even $84^{\circ}$ 
at large radii \citep{BC}.  I do not make use of data with such large disparities in their inclination estimates.  In order to
expand the size of the sample as they do, \citet{FS11} must invariably do so.  It therefore comes as no surprise that they
find a larger scatter and higher $\chi^2$, as these are the expected results of incorporating less accurate data.

Returning to the issue of the scatter, one of the brilliant aspects of the program of \citet{benfit} is that it optionally allows
one to fit for the intrinsic scatter.  Using this option returns a best fit intrinsic scatter of zero.  Zero is the preferred amount
of intrinsic scatter the program returns irrespective of whether I restrict the slope and normalization to the MOND values 
or leave them free.  Weiner's program provides no independent substantiation of the claim of finite intrinsic scatter made 
by \citet{FS11}.  

I value intellectual honesty very highly.  If I had found any serious error in my work, I would be eager to resolve the
discrepancy and say so.  Perhaps I've missed something, but I do not see what.  
Stepping back from the details, \citet{FS11} find essentially the same result as I do:  
an acceleration scale consistent with MOND with very little intrinsic scatter in the BTFR.  They just don't like it.

\section{The Real Point}

I am happy to consider \LCDM\ models that make a legitimate effort to explain the data.
\citet{gnedin11} and \citet{FS11} fall short of this standard.
Indeed, they do no even attempt to meet it.

The point is that MOND provides an economical description of galaxy scale kinematics.
The simple formula proposed by Milgrom is the \textit{effective} force law
in rotating galaxies.  This is what we need to explain, in \LCDM\ or an other theory.

It is not easy to understand how this simple formula follows from the complex physics of galaxy formation in \LCDM.
If it were, there would already be a well established model for it.  Instead, the literature is littered with mutually inconsistent
models that persistently fail to describe galaxy data as well as MOND.  

The BTFR connects the observed baryonic mass to the characteristic circular velocity.
This relation between global quantities is just one manifestation of MONDian phenomenology.
There is also a point by point mapping between the observed distribution of baryons and the observed rotation curve,
including all the bumps and wiggles in both.  This is a reiteration of the well established observational
fact that MOND fits the rotation curves of galaxies \citep{SMmond}.  

One can write this as a scaling relation $\nu$ from the rotation curve that
Newtonian gravity predicts for the observed baryonic components $V_b(r)$ to the actual observed rotation curve $V_c(r)$:
\begin{equation}
V_c^2(r) = \nu(g_N/a_0) V_b^2(r).
\label{MDaccscaling}
\end{equation}
As I've written it here \citep[see, e.g.,][]{MSclusters,M08}, $g_N = V_b^2/r$ is the Newtonian acceleration produced by
the observed stars and gas as determined by numerical solution of the Poisson equation.  There is no approximation
(e.g., exponential disks); the actual observed surface density is used, including bulge, disk, and gas, with all their bumps and
wiggles.  For reasons not presently understood, the simple, smooth function $\nu$ maps from the observed baryon distribution
to the observed rotation curve.  Only the observed baryons are required as input, with no reference to the dynamically
dominant dark matter.

Stated this way, equation~\ref{MDaccscaling} is merely a restatement of MOND.
However, the scaling relation still holds even if MOND is incorrect, as it is known to fit the majority
of rotation curves.  There is only one tiny bit of leeway.  Computing the baryonic rotation curve
\begin{equation}
V_b^2(r) = V_g^2(r) + V_*^2(r)
\end{equation}
requires knowledge of the mass-to-light ratio of the stars.  (Note that this does not apply to the gas, but one can in
principle have separate mass-to-light ratios for the bulge and disk.)  This stellar mass-to-light ratio \ML\ is the one fit 
parameter of MOND rotation curve fits.  Since we are now considering the case where such fits are just a scaling relation,
not a fundamental law of nature, there is no need for this to be the correct mass-to-light ratio.  
We can generalize the relation to account for this by defining the ratio
\begin{equation}
{\cal Q} \equiv \frac{\ML^{\mathrm{TRUE}}}{\ML^{\mathrm{MOND}}}
\end{equation}
which is the ratio of the actual stellar mass-to-light ratio to that required to obtain a MOND fit.

We can now write 
\begin{equation}
V_c^2(r) = \nu(g_N/a_0) [V_g^2(r) + V_*^2(r)/{\cal Q}].
\label{MDacc}
\end{equation}
This mass-discrepancy--acceleration relation is purely empirical and generally valid \citep{MDacc}.
It is equivalent to MOND in the case of ${\cal Q} = 1$, but it holds even when MOND does not.
This encapsulates the fact that it is usually possible to obtain a reasonable
fit to rotation curves with MOND.  The MOND mass-to-light ratios are fairly reasonable in terms of stellar populations
\citep{MDacc},
so presumably ${\cal Q}$ is not very different from unity, though it could be.  Note that this empirical formulation
need not apply to non-rotating systems; perhaps it is a scaling relation specific to disk galaxies.

While we need not explain MOND in \LCDM, we do need to explain the data.  The data for disk galaxies are 
encapsulated by equation~\ref{MDacc}.  A satisfactory model would show how this equation emerges from
the complicated process of galaxy formation.  It need not even hold all the time --- there could be some intrinsic
scatter around this relation.  But it does need to hold very nearly true most of the time if we are to achieve an
acceptable description of observational reality.

We are far from realizing this ideal.  In order to apply this test to a \LCDM\ model, said model
must make a detailed prediction for the distribution of baryons from which $V_g(r)$ and $V_*(r)$ are calculated.
Yet the distribution of baryons is notoriously difficult to predict in numerical simulations.
Until we have models that make such a prediction and actually pass this test, it seems rather sanguine to assume 
that it will all magically work out.  

Indeed, what disturbs me is the degree of fine-tuning
the observed phenomenology implies.  MOND is the \textit{effective} force law in disk galaxies.  The inverse-square
law is the force-law in the solar system.  If we concoct some theory of gravity that is different in the solar system, say an inverse-cube law,
then we would be obliged to say that the effective force law only happens to look like an inverse square law because
there is dark matter arranged \textit{just so} that it always appears to be the case.  We would reject such a hypothesis
out of hand as unacceptably fine-tuned, yet the situation is analogous in galaxies.
We know the effective force law, and must perforce arrange the dark matter \textit{just so} that we can always predict
the rotation curve from the detailed distribution of baryons alone.  

I have struggled hard to understand how MOND phenomenology can arise in \LCDM.
So far, I have failed\footnote{Few others have even tried. I am only aware of three efforts in the literature, each less
convincing than the previous.  Motivated by a preliminary version of the mass discrepancy--acceleration 
relation \citep{rutgerstalk}, \citet{vdBD} ran simulations until they found a feedback prescription that leaves intact
only those galaxies that do not deviate too much from it.  If they are correct, then all other models should use the same
feedback prescription.  They assumed purely exponential disks, so the ability of the MOND formula to fit bulges and other bumps
and wiggles remains undemonstrated.  \citet{KT02} assert that the acceleration scale of MOND follows naturally from \LCDM,
arguing that the acceleration caused by galaxy scale dark matter halos is $a_{DM} \sim cH_0$ (their equation 4).
This does nothing to demonstrate the much stricter relation of equation~\ref{MDacc}.  It also misses the essential point
that the MOND scale is connected to a critical \textit{baryonic} surface density \citep[$a_0 = G \Sigma_c$:][]{milgromstability}
for which there is at present no explanation in \LCDM.  \citet{dunkel} appears simply to present a tautology.}. 
At least I am trying to address the problem.  My critics are simply ducking it.

It is easy to believe that MOND is incorrect as a fundamental theory.  I have reviewed many dozens and perhaps hundreds of
claims to falsify it.  Most are unconvincing or obviously wrong.  Two hold water:  the residual mass discrepancy in clusters
of galaxies \citep[e.g.,][]{sanders2003,sanders2007,angusbuote} --- the bullet cluster \citep{clowe} is merely a 
dramatic example of a more general rule --- 
and the third peak of the CMB \citep{WMAPspergel}.  It might be possible to explain these in MOND, just as it may be possible
to explain MONDian phenomenology in \LCDM, but \LCDM\ is the more natural explanation for these particular observations.

Do clusters and the CMB suffice to prove the existence of non-baryonic cold dark matter?  Certainly not clusters --- there is
nothing about the missing mass in clusters that specifically requires the unseen mass to be 
non-baryonic\footnote{Apparently it needs to be collision-less in the bullet cluster; baryonic MACHOs would satisfy
that.  The collision velocity of the bullet cluster \citep{angmcg} and the morphology other bullet-like systems  
\citep[e.g., Pandora's cluster:][]{pandora} complicate the interpretation for any paradigm.}.  The third peak of the
CMB is more persuasive, as it implies the oscillation of a mass component not coupled to the baryons.  That does say
cold dark matter to me, but I cannot claim to rule out the possibility of similar behavior from some additional scalar or 
tensor field that might be introduced to combine MOND with General Relativity \citep{TeVeS,skordis}.  

Definitive proof of the existence of non-baryonic cold dark matter requires laboratory detections.
Recently, the CRESST-II collaboration \citep{CRESSTii} has claimed a detection of WIMPs.
Unfortunately, there is as yet no corroboration: their claimed detection violates limits set by 
CDMS-II \citep{CDMSii}, EDELWEISS \citep{EDELWEISS}, and XENON100 \citep{xenon100}
while being different from the claimed detection of DAMA \citep[e.g.,][]{DAMA}.
Clearly there remains a lot to sort out.

If WIMPs are detected in the laboratory, then an essential tenant of \LCDM\ is confirmed.  In this case, we still
need to explain the phenomenology encapsulated in the mass discrepancy--acceleration 
relation \citep{MDacc}.  I desperately hope that we detect WIMPs soon so that we'll know and can get on with it.  

Of course, it is dangerous to \textit{want} a particular result in science.
What if we do not detect WIMPs?  At what point do they become
untenable as dark matter candidates?  If we are forced to abandon WIMPs, what then?  Axions?  If those don't pan out,
do we just invent some other, even harder to detect particle?  If that fails, why not do it again?  And again and again, 
\textit{ad infinitum}?  While it is possible to exclude specific dark matter candidates, the concept of dark matter generally
is not falsifiable.  If it is not falsifiable, does it qualify as science?


\bibliography{btf}

\bibliographystyle{apj}

\end{document}